\documentstyle[epsfig,12pt]{article}
\begin{document}
\begin{flushright}
DTP-98/42\\
June, 1998\\
\end{flushright}
\vspace*{0.7cm}
\begin{center}
{\Large{\bf{Mass Production}}}\\
\vspace*{2mm}
{\Large{\bf{Requires}}}\\
\vspace*{2mm}
{\Large{\bf{Precision Engineering}}}
\vspace*{.7cm} \\
{\large{M.R. Pennington}}
\vspace*{0.5cm} \\
{ Centre for Particle Theory, University of Durham,
Durham DH1 3LE, U.K.}
\end{center}
\vskip 0.7cm

\begin{abstract}
\baselineskip=4.5mm
\vspace*{4mm}
This talk is in two parts both entitled {\it mass production requires
precision engineering}.  The first is about the dynamical  generation 
of mass for matter particles in gauge theories. I will explain how
the details of this depend
on a precision knowledge of the interactions.  The second is about
tests of the mechanism of chiral symmetry breaking in QCD that the precision
engineering of high luminosity colliders and particle detectors will shortly make possible.
 Since the latter topic has been described in Ref.~1, here I will just discuss the first:
the production of mass from nothing. 
\end{abstract}

\vspace*{6mm}
\baselineskip=6.5mm

\section{The problem of mass production}

  The Standard Model of the strong, weak and electromagnetic interactions is highly successful
at collating and correlating a vast amount of experimental information
in terms of a relatively simple Lagrangian. However, this involves a large number of
{\it free} parameters --- parameters that have to be fixed from experiment ---
the masses of the quarks and leptons and the CKM matrix elements that relate the mass
eigenstates to those seen by the charged weak current. In the Standard Model, these
are fixed by the couplings to the Higgs.  So if we observe the Higgs boson at the LHC, we can measure
its couplings to each fermion-antifermion pair and check that these agree
with the Standard Model.  However, even if these do agree, this won't explain {\bf why} 
they have the couplings they do.
This is determined by dynamics beyond the Standard Model,
by some interactions we have not yet discovered.  This naturally leads us to ask the
question {\it when can masses be generated dynamically in a gauge theory?}
This was asked by Miransky et al.~$^{2}$, Maskawa and Nakijima and many others~$^{3}$ 
some time ago, building on the even earlier classic papers of
Johnson, Baker and Willey~$^{4}$.
The discussion of this can be made sufficiently straightforward that one can readily
answer this question.

  We ask when can it 
be that, though the bare masses in
the Lagrangian of a gauge interaction are zero, non-zero physical
 masses are generated?  This must be a strong physics problem. 
  It is well-known that if the bare mass
is zero, then it remains zero at each order in perturbation theory.
Consequently, mass generation must be non-perturbative. 
 The lattice is often claimed to be the way
to solve such problems.  However, though the lattice nicely regulates the
ultra-violet behaviour of the interactions, it is not possible to put  massless
particles on a finite size lattice.  Consequently, lattice  calculations have to
be performed for a series of non-zero bare masses and 
then as well as taking the
lattice spacing to zero we must extrapolate to zero mass.
 Whether a dynamical mass results is all in this extrapolation, which can really only be
done if one already knows the answer. This makes the continuum the natural place
to study such a strong physics problem.

\begin{figure}[th]
\begin{center}
\mbox{~\epsfig{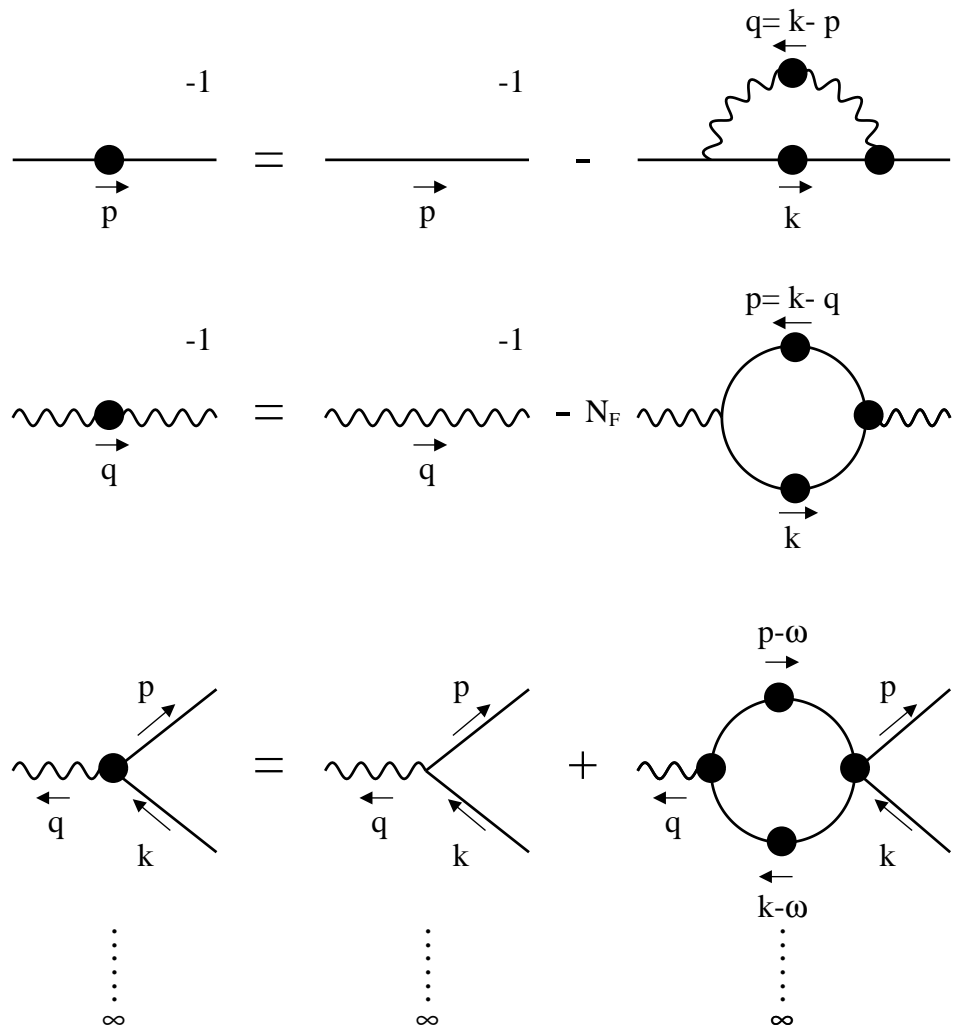}}
\end{center}
\caption[Fig.~1]{\leftskip 1cm\rightskip 1cm{Schwinger-Dyson equations for fermion, boson and vertex in QED.
The solid dots mean the Green's functions are fully dressed.}}
\end{figure}

   The field equations of a theory are the Schwinger-Dyson equations~$^{5}$.  For illustrative
purposes, let us consider these for an Abelian QED-like theory, displayed in Fig.~1. 
The first of these equations in Fig.~1, for instance relates the full
fermion propagator, $S_F(p)$, to the full photon
propagator $\Delta_{\mu\nu}(q)$ and the complete
fermion-boson vertex $\Gamma^{\mu}(k,p)$ by
\begin{eqnarray}
[iS_F(p)]^{-1}\,=\,{[iS^0_F(p)]}^{-1}\,-\,\frac{e^2}{(2\pi)^4}\,
\int d^4k (i\Gamma^{\mu}(k,p))\, iS_F(k)\, (i\gamma^{\nu})\, 
i\Delta_{\mu\nu}(q) ,
\end{eqnarray}
where $q = k-p$. Such equations, relating 2- to 3-point functions
and 3- to 4-point functions, etc., contain all the information there is about the theory.  However,
because they are an infinite set of nested integral equations, they appear to
 imply that to learn about
the 2-point function (and fermion mass generation) we have to know about 
the 3, 4, 5,..., 27,...-point functions.  Indeed, this is true, 
unless we can find some sensible truncation of this system.  The only consistent way
we know, consistent with gauge invariance and multiplicative renormalizability
at each order of truncation, is to use perturbation theory. But we know
dynamical mass generation must be non-perturbative.

\baselineskip=6.5mm
\section{Quenched QED in the rainbow approximation}

  To solve the Schwinger-Dyson equations non-perturbatively,
let us proceed by crudely butchering these equations to make the problem
tractable and only worry afterwards about what we have done --- that will lead us
to precision engineering.  To treat the problem at its simplest, let us study the
fermion propagator and cut it away from all the rest of the infinite hierarchy,
cf.~Fig.~1,
by (i) considering quenched QED, so that the gauge boson propagator is bare
(i.e. set $N_F=0$),
and (ii) treating the interaction as bare. This is what is called the
{\it rainbow} approximation.  In this approximation,
 only the fermion propagator, $S_F(p)$,  is non-perturbative. Because of its spin and Lorentz structure, $S_F(p)$
 depends on two independent scalar functions. This can be expressed in several equivalent ways.
 Convenient here is to introduce the fermion wavefunction renormalization 
 ${\cal F}(p^2)$ and the mass function ${\cal M}(p^2)$, so
 $S_F(p)\,=\,{\cal F}(p^2)/(\rlap /p - {\cal M}(p^2))$.
 The bare propagator then has ${\cal F}(p^2)=1$ and ${\cal M}(p^2)=m_0$. $m_0$ is what
 we will set equal to zero later. The bare photon propagator carrying momentum $q$ is
 \begin{eqnarray}
 \Delta_{\mu\nu}\,=\,-\,\left[\frac{1}{q^2}\,\left(g_{\mu\nu}\,-\,\frac{q_{\mu}q_{\nu}}{q^2}\right)\,
 +\,\xi\frac{q_{\mu}q_{\nu}}{q^4}\right]\; ,
 \end{eqnarray}
 where $\xi$ is the usual covariant gauge parameter.  Working in 
 Euclidean space,
  the angular integrals can be done, giving two coupled equations
\begin{eqnarray}
\frac{{\cal M}(p^2)}{{\cal F}(p^2)}\,&=&\,m_0\,+\,\frac{\alpha_0}{4\pi} (3+\xi)\,\int
_0^{\Lambda^2}\, dk^2\, \frac{{\cal F}(k^2)\, {\cal M}(k^2)}{k^2+{\cal M}^2(k^2)}\,
\left[ \theta_+ \frac{k^2}{p^2}\,+\,\theta_- \right]\, ,\\\nonumber\\  
\frac{1}{{\cal F}(p^2)}\,&=&\;1\;+\;\frac{\alpha_0 \xi}{4\pi} \,\int
_0^{\Lambda^2}\, dk^2\, \frac{{\cal F}(k^2)}{k^2+{\cal M}^2(k^2)}\,
\left[ \theta_+ \frac{k^4}{p^4}\,+\,\theta_- \right]\, ,
\end{eqnarray}
where $\theta_{\pm}$ denote Heaviside step functions depending on the sign of
$(p^2-k^2)$  and
  where the coupling $\alpha_0\equiv e^2/4\pi$ --- the subscript $0$ is to
  emphasise this is quenched QED and the coupling does not run. 
  Here $\Lambda$ is an ultraviolet cut-off, introduced to make integrals finite. 
  From Eqs.~(3,4) we see we can simplify
   this coupled system further by working in the Landau gauge $\xi=0$. Then 
    by our quenched rainbow approximation, we have
   reduced an infinite set of Schwinger-Dyson equations
   to ${\cal F}(p^2)=1$ and
   \begin{eqnarray} 
   {\cal M}(p^2)\, =\, m_0\, +\, \frac{3\alpha_0}{4\pi}\, \int
_0^{p^2}\ \frac{dk^2}{p^2} \frac{k^2 {\cal M}(k^2)}{k^2+{\cal M}^2(k^2)}
+\, \frac{3\alpha_0}{4\pi}\, \int
_{p^2}^{\Lambda^2}\  {dk^2}\  \frac{ {\cal M}(k^2)}{k^2+{\cal M}^2(k^2)} . 
\end{eqnarray}
   
    Now what we want to know is, if $m_0=0$, when can ${\cal M}(p^2)$ be 
   non-zero? We see that with $m_0=0$, Eq.~(5) has a solution
   ${\cal M}(p^2)=0$.  This always happens if the interaction involves an odd
   number of gamma matrices (as the assumed bare one does here).  To see
   that ${\cal M}(p^2)$ can be non-zero, let us convert this equation
   into a differential one, giving
   \begin{eqnarray} 
   {d\over dp^2}\,\left(p^4 {d\over dp^2}\,{\cal M}(p^2)\right)\,=\,-\,
   {3\alpha_0\over{4\pi}}\,{p^2 {\cal M}(p^2)\over{p^2+{\cal M}^2(p^2)}}\; . 
    \end{eqnarray}
Notice
    that at large momenta, when $p^2\gg {\cal M}^2(p^2)$, this equation linearises
    and so then has the solution ${\cal M}(p^2) = A (p^2)^s$.
    With this, Eq.~(6) gives
    $s(s+1) = -3\alpha_0/4\pi$, which has the solutions
    $s=-\frac{1}{2} \pm \frac{1}{2} \sqrt{1-\frac{3\alpha_0}{\pi}}$.  We see
    that the character of the solutions differ depending on whether
    $\alpha_0$ is greater or less than $\pi/3$.  If $\alpha_0 < \pi/3$, the solutions have simple
    power behaviour at large momenta, but if $\alpha_0 > \pi/3$, then the solutions are oscillatory
    behaving like
\begin{eqnarray}
\nonumber 
    \frac{A}{p}\,\sin\left(\frac{p^2}{2} \sqrt{ \frac{3\alpha_0}{\pi} -1} 
    + \delta\right)\; .
    \end{eqnarray}
\noindent Now when an integral equation is converted to a differential
    equation, we lose information about integration constants.
    In particular, we have the boundary condition 
    \begin{eqnarray} 
    \lim_{p^2\to\Lambda^2}\,\left\{{\cal M}(p^2)\,+\,p^2\, \frac{d}{dp^2}\,
    {\cal M}(p^2)\right\}\,=\,m_0\; .
    \end{eqnarray}
If $m_0=0$, it is only the oscillating behaviour that can satisfy this boundary condition.
Thus a fermion mass can be dynamically generated provided the interaction is strong enough, i.e.
$\alpha_0 \ge \pi/3$.

\begin{figure}[th]
\begin{center}
\mbox{~\epsfig{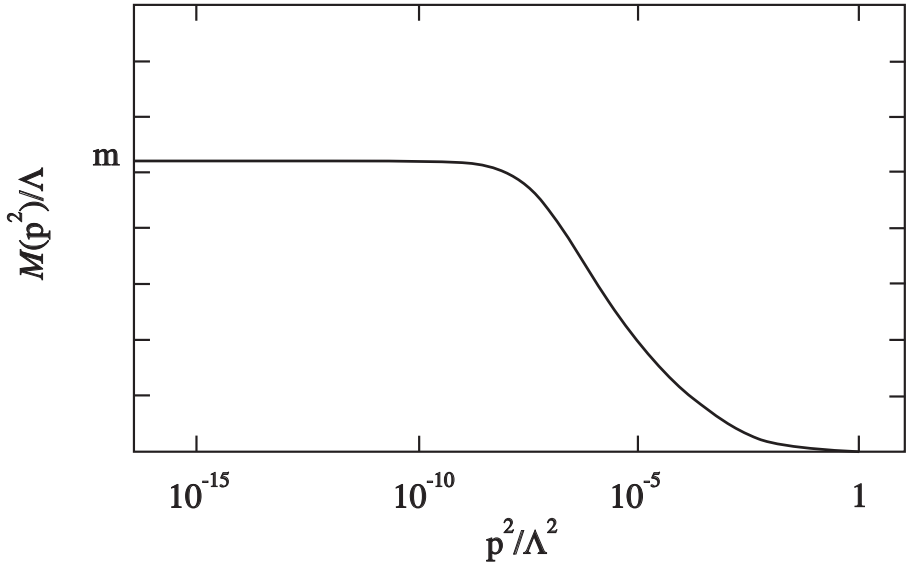}}
\end{center}
\vspace*{-1mm}
\caption[Fig.~2]{\leftskip 1cm\rightskip 1cm{The mass function, ${\cal M}(p^2)$, for some
coupling $\alpha_0$ above its critical value.}}
\vspace*{-3mm}
\end{figure}

  It is useful to look at the function ${\cal M}(p^2)$ as a
 function of $p^2$, Fig.~2,
at some $\alpha_0 > \pi/3$ found by solving Eq.~(5).  This has a shape
characteristic of all mass functions.  It is flattish with a value we may call $m$ at low
spacelike momenta, where $\mid p^2 \mid < m^2$, and then 
when $\mid p^2 \mid \gg m^2$  it has a power fall off,
with oscillations beyond the ultaviolet cut-off.  The {\it Euclidean} mass
$m$ defined either as ${\cal M}(p^2) = p = m$ or simply as ${\cal M}(0) = m$
(these $m$'s are qualitatively similar) can be taken as a guide to the size of the mass that is
generated and so parametrizes the scale and shape of the mass function.  We can calculate $m$ as a function of $\alpha$
numerically and see (Fig.~3) that for $\alpha_0 < \pi/3$, there is only
the massless solution, but when $\alpha_0$ becomes bigger than $\pi/3$
 a second solution
bifurcates away from the massless one with an analytic form just as
calculated by Miransky~$^{2}$.
\begin{figure}[bht]
\begin{center}
\mbox{~\epsfig{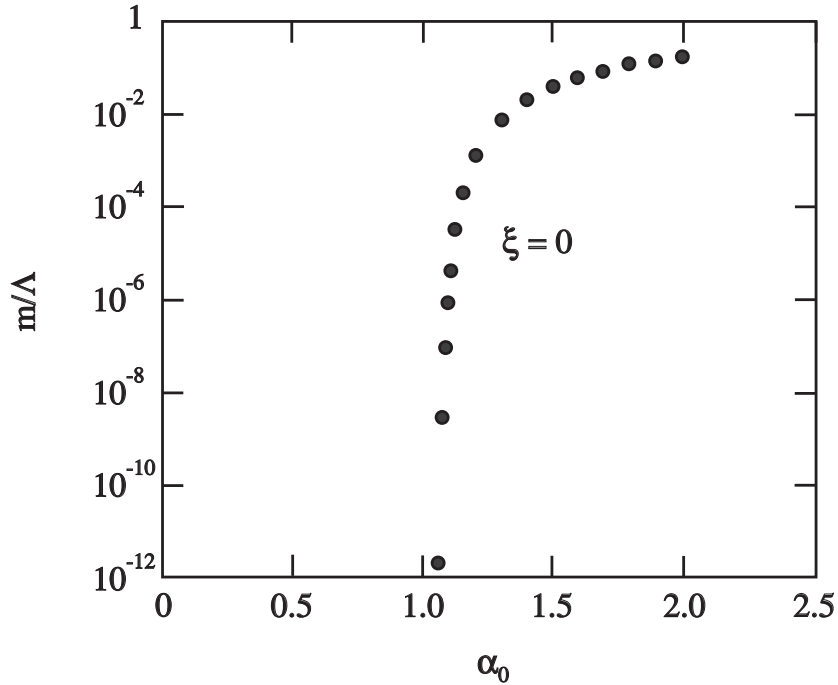}}
\end{center}
\vspace*{-1mm}
\caption[Fig.~3]{\leftskip 1cm\rightskip 1cm{The Euclidean mass $m$ 
as a function of the coupling $\alpha_0$ in the {\it rainbow} approximation
 in the Landau gauge.
$\Lambda$ is the ultraviolet cut-off.}}
\end{figure}

So far we have  calculated this mass in just one gauge, the Landau gauge.  But this mass is,
in principle, an observable: imagine a world with quenched QED and massless electrons.  Such an electron would propagate
at the speed of light, but if it came close to a heavy nucleus, where the effective
coupling $Z\alpha$ became greater than $\pi/3$, such an electron would have a mass generated
and so would no longer move at the speed of light. That this massive solution is
energetically favoured over the massless one can be shown by considering the
Cornwall-Jackiw-Tomboulis potential~$^6$.  The critical value of the coupling is an observable and hence gauge independent
in this world, so let's compute this in some other gauge, by solving the
 coupled system, Eqs.~(3,4) with $\xi\ne 0$.
Fig.~4 shows the mass $m$ for $\xi=1$ and 3 to compare with the Landau gauge
result. While qualitatively similar with a non-zero mass generated if the coupling is large enough, the
critical coupling, and hence the mass that is generated, is strongly
gauge dependent.  This is surely worrying for a supposedly physical quantity.

\begin{figure}[th]
\begin{center}
\mbox{~\epsfig{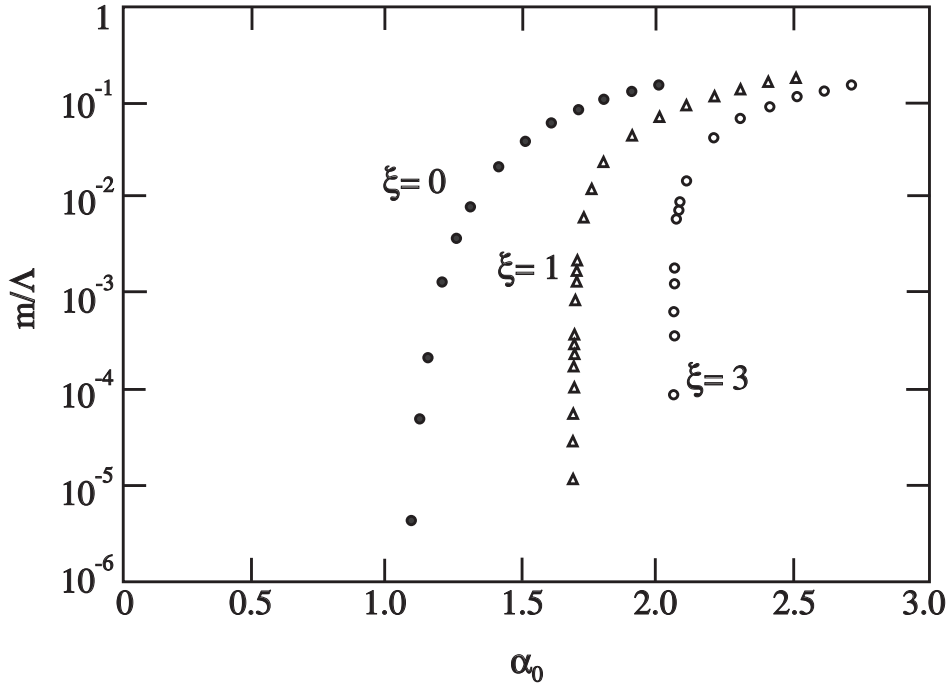}}
\end{center}
\vspace*{-1mm}
\caption[Fig.~4]{\leftskip 0.85cm\rightskip 0.85cm{The Euclidean mass $m$ as a 
function of the coupling $\alpha_0$ in the {\it rainbow} approximation in 
different covariant gauges.
$\Lambda$ is the ultraviolet cut-off.}}
\end{figure}

\baselineskip=6.25mm

\section{Quenched QED -- towards a consistent truncation}
   That this has happened should not come as a total surprise, since clearly our crude approximation
of using a bare interaction violates an important consequence of
gauge invariance, namely the Ward and Green-Takahashi identities~$^7$.  How to solve these
was set out by Ball and Chiu~$^8$  giving the {\it longitudinal part} 
\begin{eqnarray}
\nonumber \Gamma^{\mu}_L(k,p)&=&  \frac{1}{2}\left( \frac{1}{{\cal F}(k^2)}\ +
  \frac{1}{{\cal F}(p^2)}\  \right)\,\gamma^{\mu}\,
  - \,\mbox{}   \left(
  \frac{{\cal M}(k^2)}{{\cal F}(k^2)} - \frac{{\cal M}(p^2)}{{\cal F}(p^2)}\ \right)
\frac{(k+p)^{\mu}}{k^2 - p^2}\,
\\
&&+ \mbox{}  \frac{1}{2}\ \left(
  \frac{1}{{\cal F}(k^2)} - \frac{1}{{\cal F}(p^2)}\ \right)
\frac{1}{k^2 - p^2}\,(\not \! k +\not \! p)
               (k+p)^{\mu}
                \, ,
\end{eqnarray}
where $k$ and $p$ are the fermion momenta. Notice that this tells us that a key part of the 3-point interaction is
determined by the fermion propagator. 
To this can be added any  {\it transverse} part, $\Gamma^{\mu}_T(k,p)$,
specified by $q_{\mu}\ \Gamma_T^{\mu}(k,p) = 0$ and  which must
satisfy $\Gamma_T^{\mu}(p,p) = 0$, so that $\Gamma^{\mu}_L(k,p)$ of Eq.~(8) alone
fulfills the Ward identities. Now it is well-known that the vector-fermion-antifermion
coupling involves 12 independent vectors.  However, one has zero coefficient because the
Ward-Green-Takahashi identity has no $\sigma^{\mu\nu}$ component.  This leaves 11
vectors. Three appear in the Ball-Chiu longitudinal part, Eq.~(8),
 and so eight are transverse, the $T^{\mu}_i(k,p)$,
 to the boson momentum, $q = k - p$. 
 As an example $T^{\mu}_6(k,p) = \gamma^{\mu} (k^2-p^2) - (k+p)^{\mu}(\not \! k + \not \! p)$.
  Thus the transverse component can be written as 
 \begin{eqnarray} 
  \Gamma^{\mu}_T(k,p)\,=\,\sum_{i=1}^8 \tau_{i}(k^2,p^2,q^2)\ T_{i}^
     {\mu}(k,p)\, .
     \end{eqnarray}
 Since
we appear to know nothing about these components, we first set the
coefficients $\tau_i$ to zero and use just the Ball-Chiu longitudinal
vertex.  Solving the equivalent of Eqs.~(3,4), one finds  dynamical mass
generation is still gauge dependent.  This is because to deal with these equations, 
 they have to be regulated consistently.  This means the fermion
propagator must be multiplicatively renormalizable. 
This is implicit, for instance, in the proof by Atkinson and Fry~$^9$ 
that the position of the pole in the
propagator is gauge independent. Consequently,
 multiplicative renormalizability  must require
the transverse part of the vertex to be non-zero. Like the 
longitudinal part, Eq.~(8), it must be determined (at least in part) by the
fermion propagator.  This suggests that multiplicatively renormalizable might, together with gauge invariance
and gauge covariance, determine the nature of the interaction, at least as far as the
Schwinger-Dyson equation for the 2-point function is concerned.

\begin{figure}[th]
\begin{center}
\mbox{~\epsfig{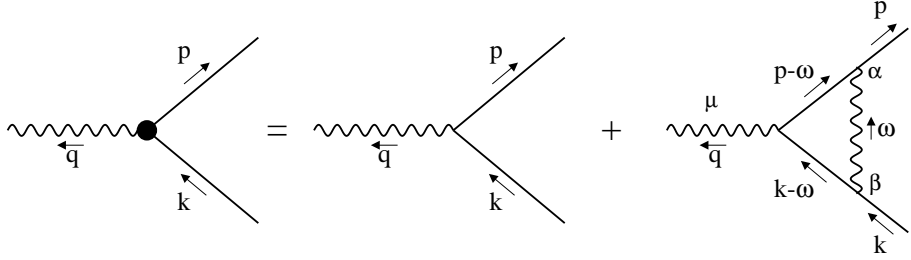}}
\end{center}
\caption[Fig.~5]{\leftskip 1cm\rightskip 1cm{One loop 
perturbative corrections to the fermion-boson vertex, $\Gamma^{\mu}(k,p)$.}}
\end{figure}

  Since these constraints are, of course, satisfied order by order in perturbation theory, we
might look at the perturbative result for $\Gamma^{\mu}(k,p)$, where $k,p$ are
the fermion momenta, Fig.~5. At zeroth order, this is just $\gamma^{\mu}$, and is
given by the Ball-Chiu vertex with ${\cal F}=1$ and ${\cal M}=$ const.
This tells us that the transverse vertex and hence the $\tau_i$ of Eq.~(9)
must be  ${\cal O}(\alpha)$ in perturbation theory. What the $\tau_i$ are to
${\cal O}(\alpha)$ has in fact been calculated (Fig.~5) in an arbitrary covariant gauge
relatively recently by Ay\c se K{\i}z{\i}lers\"{u}, 
Manuel Reenders and myself~$^{10}$. Indeed, it is only at this
meeting that Ay\c se and I have met Manuel face-to-face instead of over the Net. 
The result for each of the eight $\tau_i$ is so complicated that it is difficult to recognise a simple pattern. However,
multiplicative renormalizability of the fermion propagator is related to the ultraviolet behaviour of the
loop integral.  Then the full vertex $\Gamma^{\mu}(k,p)$ is only needed in the limit
$k^2 \gg p^2$, Figs.~1,5. In this limit, as noted by Curtis and I~$^{11}$, the transverse component
is simple 
\begin{eqnarray}
{\Gamma^{\rm{(pert)}}_T}^{\mu}(k,p)\;=\,\frac{\alpha_0 \xi}{8 \pi  k^2}\ 
\ln \frac{k^2}{p^2}\,\left(\gamma^{\mu}-\frac{k^{\mu}k}{k^2}\right)\; . 
\end{eqnarray}
This can be expressed at ${\cal O}(\alpha)$
in a way related to fermion functions and just the vector $T_6^{\mu}$: 
\begin{eqnarray}
\Gamma^{\mu}_T(k,p)&=&-\,\frac{1}{2}\ \frac{k^2+p^2}{(k^2-p^2)^2}
    \left( \frac{1}{{\cal F}(k^2)} - \frac{1}{{\cal F}(p^2)} \right)\,T_6^{\mu}(k,p)\, .
\end{eqnarray}
This, when added to the Ball-Chiu longitudinal part of Eq.~(8), gives what  is known as the CP vertex.

  The Ward-Green-Takahashi identity requires that the vertex cannot just
be some factor times the bare vertex $\gamma^{\mu}$. It must involve some of the other 10 vectors too.
Moreover, multiplicative renormalizability tells us that the coefficients of these vectors must
involve the inverse of the fermion wavefunction renormalization, ${\cal F}$,
just as in Eqs.~(8,11). This, of course, has not stopped ansatze for the vertex,
like $\gamma^{\mu}/({\cal F}(p^2) {\cal F}(k^2))$, which cannot be sensible in
4, or even 3 dimensions, being proposed.

  To understand in general how multiplicative renormalizability imposes constraints on the transverse vertex,
let us consider a perturbative expansion of the solution to the
Schwinger-Dyson equation for ${\cal F}(p^2)$, just as an illustration.  We would find in
a leading logarithmic expansion
\begin{eqnarray}
\nonumber 
{\cal F}(p^2)\, =\, 1\, +\, \alpha_0 A_1 \ln p^2/\Lambda^2 &+&
\alpha_0^2 A_2 \ln^2 p^2/\Lambda^2\, +\, \cdots\,+\,
\alpha_0^n A_n \ln^n p^2/\Lambda^2\, +\, \cdots\; . 
\end{eqnarray}
The $A_n$ will be related to $A_1, ..., A_{n-1}$ and the parameters of the transverse vertex.
However, in massless quenched QED, multiplicative renormalizability requires $A_2=A_1^2/2$ (and $A_n=A_1^n/n!$
in general).  In fact, the leading logs must exponentiate to give 
${\cal F}(p^2) = (p^2/\Lambda^2)^{\nu}$, where $\nu = \alpha_0 A_1$ and
explicit calculation gives $A_1=\xi/4\pi$. Indeed, 
\begin{eqnarray}
{\cal F}(p^2) = (p^2/\Lambda^2)^{\gamma}
\end{eqnarray}
 is the only multiplicatively renormalizable form
in a one scale problem, where  $\gamma-\nu$ is not only
${\cal O}(\alpha_0^2)$, but  gauge independent too~$^{12}$.
The CP form, Eqs.~(8,11), reproduces this.
Being able to relate the $A_n$ fixes the transverse vertex or at least its leading
log expansion.  If we use the transverse form suggested by CP, then the mass generation
becomes almost independent of the gauge~$^{13}$ as it must, Fig.~6.

\begin{figure}[th]
\begin{center}
\mbox{~\epsfig{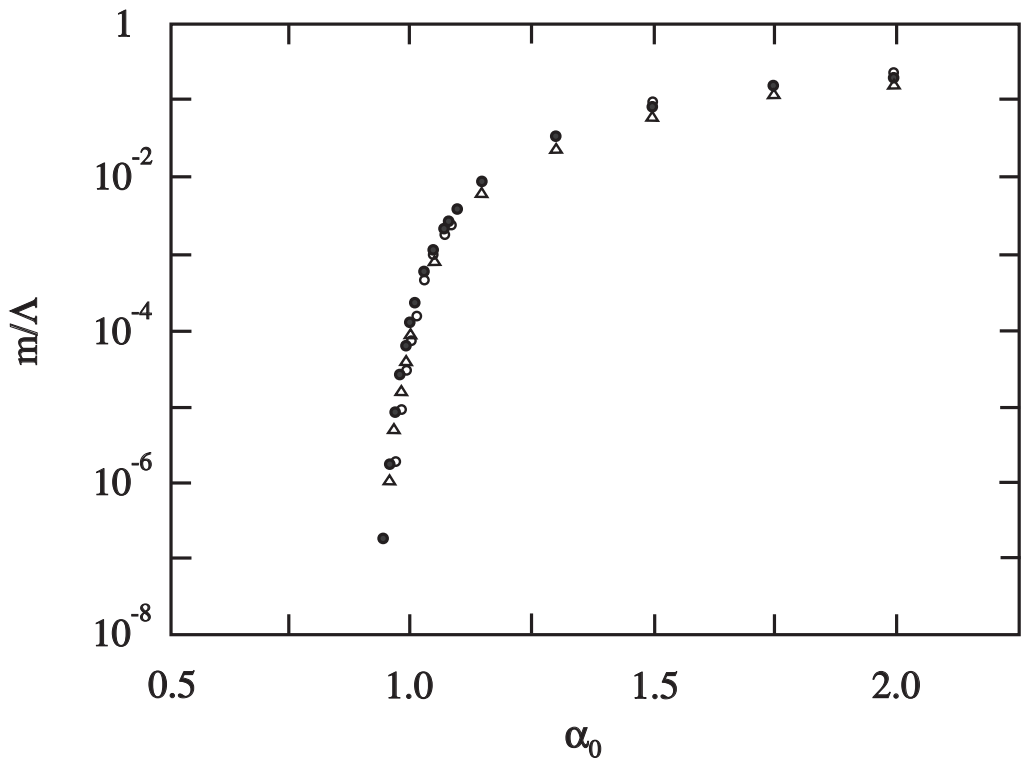}}
\end{center}
\caption[Fig.~6]{\leftskip 1cm\rightskip 1cm{The Euclidean mass $m$ as a 
function of the coupling $\alpha_0$ with the CP vertex
in the same covariant gauges as Fig.~4.
$\Lambda$ is the ultraviolet cut-off.}}
\end{figure}

  In the simple massless quenched situation
we consider here, one readily sees that the power behaviour of
the fermion wavefunction renormalization, Eq.~(13), results from a specific part
 of the integral of Eq.~(1)
and hence the rest must be zero.  This was first noted by Burden and Roberts~$^{14}$,
and recently referred to as the {\it transverse condition}~$^{14}$.
How to construct a solution of this condition was first discussed by Dong et al.~$^{15}$
and some of their simplifying assumptions have been relaxed by Bashir et al.~$^{16}$.
The solution is fixed up to some unknown, but constrained, function
$U_1(x)$, where for instance
\begin{eqnarray}
\nonumber
\tau_{6}^{\rm{eff}}(k^2,p^2)&&=-\,\frac{1}{2}\ \frac{k^2+p^2}{(k^2-p^2)^2}
    \left( \frac{1}{{\cal F}(k^2)} - \frac{1}{{\cal F}(p^2)} \right)\\ \nonumber
  &&+ \frac{1}{12}\ \frac{1}{s_{1}(k^2,p^2)}
     \Bigg[ \frac{3k^2-p^2}{(k^2-p^2)^2}\, U_{1}\left(\frac{k^2}{p^2}\right)  +
    \frac{k^2-3p^2}{(k^2-p^2)^2}\, U_{1}\left(\frac{p^2}{k^2}\right)  \Bigg]\\
 &&  - \frac{2\pi}{3\alpha}\ \frac{\gamma-\nu}{(k^2-p^2)^2}
    \left[ \frac{3k^2-p^2}{{\cal F}(k^2)} + 
    \frac{k^2-3p^2}{{\cal F}(p^2)} \right]\, ,
\end{eqnarray}
\noindent with 
\begin{eqnarray*}
 s_{1}(k^2,p^2)= \frac{k^2}{p^2}\ {\cal F}(k^2) +
\frac{p^2}{k^2}\ {\cal F}(p^2)\, ,
\end{eqnarray*}
and $\tau_6^{\rm{eff}}(k^2,p^2)$ is the angular average of $\tau_6(k^2,p^2,q^2)$
defined in Ref.~16.
\noindent This coincides with the CP result, Eq.~(11), if $U_1 \to 0$
in the limit $k^2 \gg p^2$.
 However, non-zero $U_1$ is essential when $k^2 \to p^2$ to remove the 
 potential kinematic singularity in the CP form, Eq.~(11), as noted 
 in Refs.~14,15.

\section{Full QED and the future}
 
   Now we want to extend this to unquenched QED, where the two scale nature of the problem
 (essentially $p$ and $\Lambda_{QED}$) means that the non-perturbative form
 for ${\cal F}(p^2)$ cannot be found from the renormalization group without
 a complete knowledge of the (non-perturbative) $\beta$--function.  The solution
 even in the leading logarithmic approximation is a very lengthy and painful calculation,
 which Ay\c se K{\i}z{\i}lers\"{u} has worked out~$^{17}$.  The idea is to construct a full vertex that ensures
 the multiplicative renormalizability of both the fermion and
  photon propagators, Fig.~1.  These
 are, of course, strongly coupled.  Moreover, the ultraviolet behaviour of the
  loop corrections to each explore
 quite distinct kinematic regimes of $\Gamma^{\mu}(k,p)$. As already remarked,
 for the fermion propagator, this is $k^2, q^2 \gg p^2$, while for the photon
 propagator, this is $k^2, p^2 \gg q^2$ --- see Fig.~1.  In ${\cal O}(\alpha)$
 perturbation theory, the vertex has quite different behaviours in these limits.
 While the former (fermion) limit gives the $\ln k^2/p^2$ factors for $\tau_6$
 in Eq.~(10), the latter (photon) limit gives factors of $\ln q^2/k^2$ for
 $\tau_2,\ \tau_3$. If these ${\cal O}(\alpha)$ perturbative results are
 to be expressible in terms of wavefunction renormalizations, then we must
 not only have structures like
\begin{eqnarray*}
 \left( \frac{1}{{\cal F}(k^2)} - \frac{1}{{\cal F}(p^2)} \right)
 && {\rm seen~~in~~Eq.~(11),}\\ &&{\rm  but~~also}\quad
 \left( \frac{1}{{\cal F}(k^2)} + \frac{1}{{\cal F}(p^2)} -
 \frac{2}{{\cal F}(q^2)}\right)\; .
\end{eqnarray*}
 During the course of this workshop, Ay\c se K{\i}z{\i}lers\"{u} and I hope to 
 finish solving these constraints and hence deduce an ansatz for the full fermion-boson vertex that leads
 to fermion and photon propagators that are multiplicatively renormalizable.

  Such a vertex is not only relevant to strong coupling
Abelian gauge theories, but to QED itself, since it may efficiently sum
key parts of the infinity of Feynman graphs needed for
gauge invariance and multiplicative renormalizability.  Of course, QED is not the end.
The eventual aim is to extend such studies to QCD to learn
how a consistent truncation procedure for non-Abelian theories can be
developed.  This is essential, if we are to understand the continuum infrared
behaviour of gluons, ghosts and quarks in a consistent way.
It is believed that the {\it up} and {\it down} quark mass functions
exhibit dynamical chiral symmetry breaking.
How the structure of the QCD vacuum generates this breaking will be tested
in present and future experiments --- in E865 at BNL within months, and
in KLOE at LNF and Dirac at CERN next year (see Ref.~1).  That is
the subject of the rest of this talk. A subject I have already summarised~$^1$.

\section*{Acknowledgments}

  This meeting was sponsored and hosted by 
the Centre for Subatomic Structure of the University of Adelaide.
My thanks are due to Tony Thomas, whose directing of the CSSM, has promoted
such valuable meetings and to Andreas Schreiber and Tony Williams for organising
this most enjoyable, stimulating and instructive workshop.

\end{document}